%% file: conference_101719.tex
\documentclass[conference]{IEEEtran}
\IEEEoverridecommandlockouts
\usepackage{cite}
\usepackage{amsmath,amssymb,amsfonts}
\usepackage{algorithmic}
\usepackage{graphicx}
\usepackage{hhline}
\usepackage{tabularx}
\usepackage{multirow}
\usepackage{booktabs}
\usepackage{textcomp}
\usepackage{xcolor}
\usepackage{tikz}
\usepackage{multirow}
\usepackage{graphicx}
\usepackage{ifthen}
\usepackage{xcolor}
\usepackage{pgfplots}
\usetikzlibrary{matrix,calc}
\usepackage{tikzsymbols}
\usepackage{tikz}
\usetikzlibrary{shapes}
\usetikzlibrary{calc, arrows, quotes}
\usepackage{pgfplots}
\pgfplotsset{compat=1.13, axis line style=thick}
\tikzstyle{arrow} = [thick,->,>=stealth]
\pgfplotsset{
  compat=newest,
  xlabel near ticks,
  ylabel near ticks
}
\usetikzlibrary{trees}
\usepackage{scalerel}
\usepackage{tikz-qtree}
\usepackage{lipsum,adjustbox}
\usepackage{url}
\usepackage[colorlinks,urlcolor=blue,linkcolor=blue,citecolor=black]{hyperref}
\usepackage[numbers]{natbib}
\usepackage{subfig}
\usepackage{svg}
\usepackage{mathtools}
\DeclarePairedDelimiter\ceil{\lceil}{\rceil}
\DeclarePairedDelimiter\floor{\lfloor}{\rfloor}
\DeclareMathOperator*{\concat}{\scalerel*{\Vert}{\sum}}
\usetikzlibrary{arrows, automata}
\usepackage{academicons}
\usepackage{orcidlink}
\definecolor{orcidlogocol}{HTML}{A6CE39}
\usepackage{moresize}
\usepackage{array}
\newcommand{\PreserveBackslash}[1]{\let\temp=\\#1\let\\=\temp}
\newcolumntype{C}[1]{>{\PreserveBackslash\centering}p{#1}}
\newcolumntype{R}[1]{>{\PreserveBackslash\raggedleft}p{#1}}
\newcolumntype{L}[1]{>{\PreserveBackslash\raggedright}p{#1}}
\usepackage{siunitx}
\usepackage{float}
\usepackage[activate=false,kerning]{microtype}
\SetExtraKerning[unit=character]{encoding=*}{\textemdash={100,100}}

\usetikzlibrary{arrows, automata}
\def\BibTeX{{\rm B\kern-.05em{\sc i\kern-.025em b}\kern-.08em
    T\kern-.1667em\lower.7ex\hbox{E}\kern-.125emX}}
\usepackage{amsthm}
\newtheorem{thm}{Theorem}[section]
\theoremstyle{definition}

\newtheorem{prop}[thm]{Proposition}

\usepackage{dsfont}
\usepackage[utf8]{inputenc}
\newcommand{\screedsolo}{\textsc{SCReedSolo}}
\newcommand{\screedsolos}{\textsc{SCReedSolo} }
\begin{document}

\title{\textsc{SCReedSolo}: A Secure and Robust LSB Image Steganography Framework with Randomized Symmetric Encryption and Reed–Solomon Coding\\
}

\author{\IEEEauthorblockN{Syed Rifat Raiyan}
\IEEEauthorblockA{
\textit{Systems and Software Lab (SSL)} \\
\textit{Department of Computer Science and Engineering} \\
\textit{Islamic University of Technology}\\
Dhaka, Bangladesh \\
rifatraiyan@iut-dhaka.edu}
\and
\IEEEauthorblockN{Md. Hasanul Kabir}
\IEEEauthorblockA{
\textit{Computer Vision Lab (CVLab)} \\
\textit{Department of Computer Science and Engineering} \\
\textit{Islamic University of Technology}\\
Dhaka, Bangladesh \\
hasanul@iut-dhaka.edu}
}

\maketitle

\begin{abstract}
Image steganography is an information-hiding technique that involves the surreptitious concealment of covert informational content within digital images. In this paper, we introduce \textsc{SCReedSolo}, a novel framework for concealing arbitrary binary data within images. Our approach synergistically leverages Random Shuffling, Fernet Symmetric Encryption, and Reed–Solomon Error Correction Codes to encode the secret payload, which is then discretely embedded into the carrier image using LSB (Least Significant Bit) Steganography. The combination of these methods addresses the vulnerability vectors of both security and resilience against bit-level corruption in the resultant stego-images. We show that our framework achieves a data payload of 3 bits per pixel for an RGB image, and mathematically assess the probability of successful transmission for the amalgamated $n$ message bits and $k$ error correction bits. Additionally, we find that \screedsolos yields good results upon being evaluated with multiple performance metrics, successfully eludes detection by various passive steganalysis tools, and is immune to simple active steganalysis attacks. Our code and data are available at \texttt{\url{https://github.com/Starscream-11813/SCReedSolo-Steganography}}.
\end{abstract}


\input{sections/1_Introduction}
\input{sections/2_Problem_Formulation}
\input{sections/4_Methodology}
\input{sections/5_Experiment}
\input{sections/6_Conclusion_and_Future_Work}
\fontsize{10}{11.5}\selectfont
\bibliographystyle{IEEEtranN}
    \bibliography{ref}

\end{document}

%% file: sections/1_Introduction.tex
\section{Introduction}
\begin{figure}[t]
    \centering
    \includegraphics[width=0.985\linewidth]{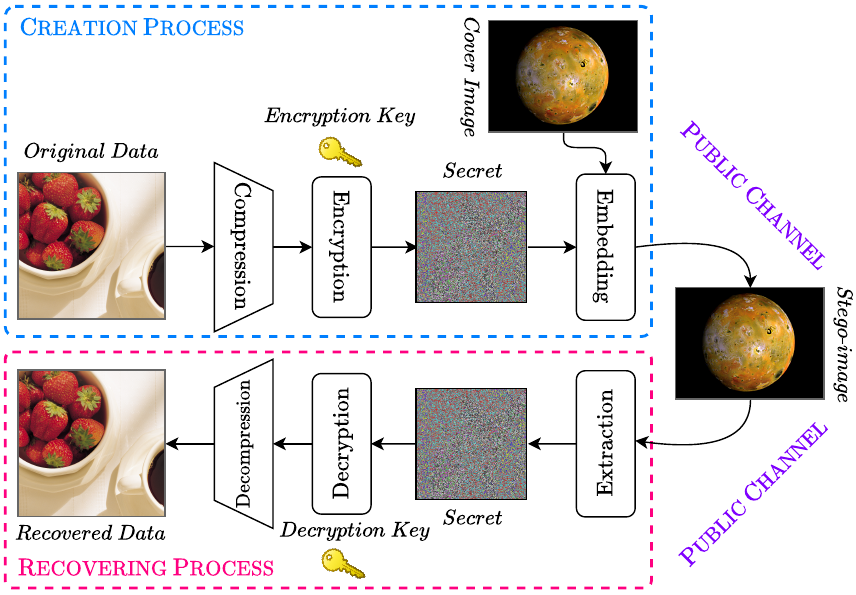}
    \setlength{\belowcaptionskip}{-18pt}
    \caption{General pipeline of a secure image steganography algorithm.}
    \label{fig:fig1}
\end{figure}

The fundamental aim of image steganography is to embed a confidential message within an image with such precision and subtlety that its presence remains wholly indiscernible to both scrutiny and suspicion. Unlike cryptographic techniques, which mainly focus on rendering messages unintelligible to unauthorized interlocutors, steganography adopts an orthogonal objective: to veil the message's very presence and perceptual detectability \cite{wang2004cyber}. In a typical use case, a sender embeds the hidden message utilizing a cover image as a substrate, which is then dispatched to the recipient through a communication channel, ostensibly indistinguishable from its unaltered counterpart to unauthorized observers. The recipient, in order to retrieve the latent message, invokes an extraction protocol---often predicated on shared cryptographic primitives or stego-key synchronization, thereby ensuring that any intercepting party remains oblivious of the message's existence \cite{lou2002steganographic}. The general pipeline of image steganography is as portrayed in Figure \ref{fig:fig1}. Steganography has been practiced for centuries, with one of its most widely recognized forms being invisible writing, often achieved through the use of invisible ink. This technique gained particular prominence during World War II \cite{kumar2010steganography}.
\begin{figure*}
    \centering
    \begin{minipage}[b][][b]{0.8\textwidth}
    \centering
        \includegraphics[width=1\linewidth]{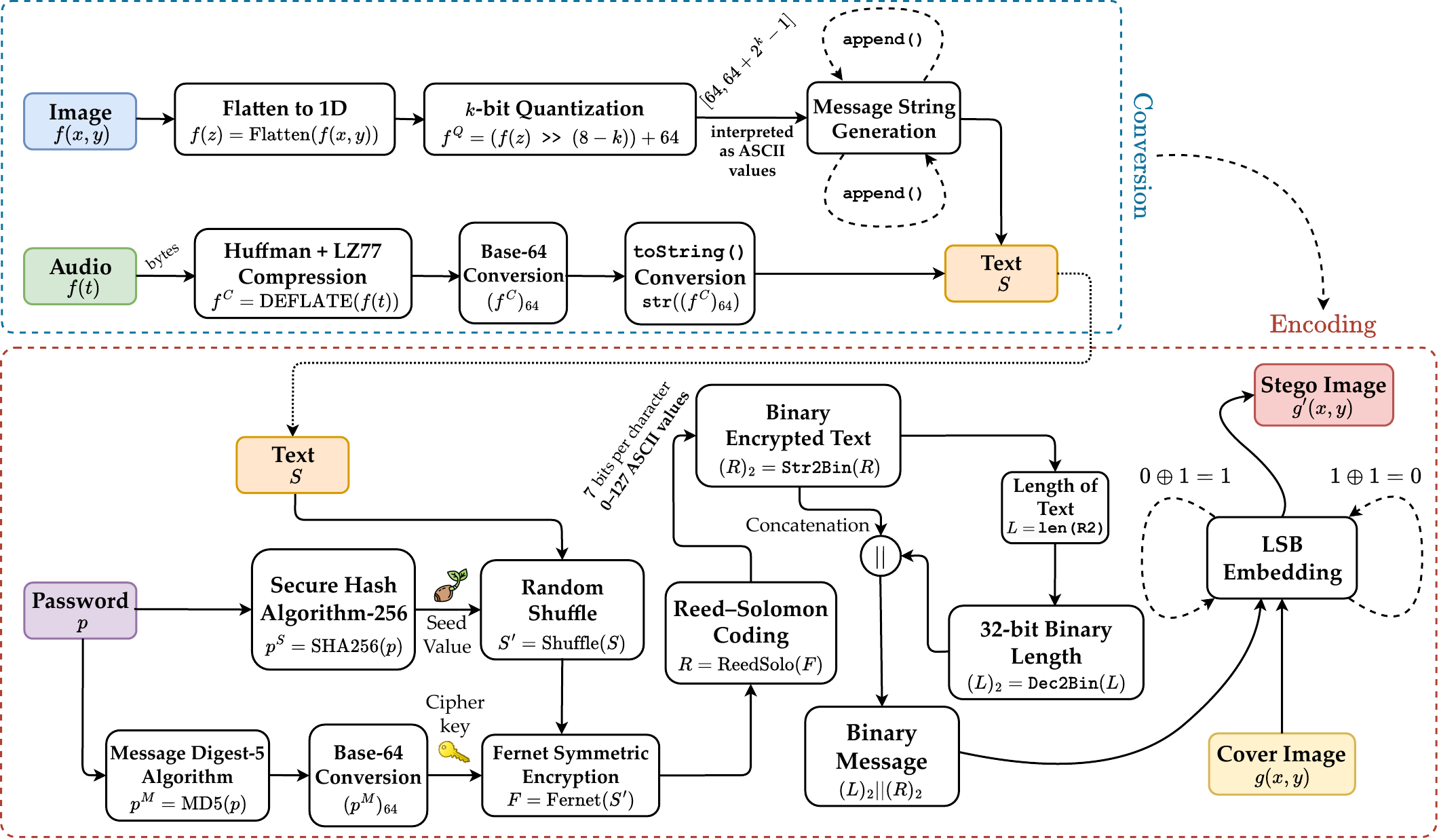}
        \setlength{\belowcaptionskip}{-15pt}
        \caption{Overview of our proposed framework \screedsolo. The workflow's upper part converts the image and audio to text, and the part beneath portrays \screedsolo's encoding process.}
        \label{fig:fig6}
    \end{minipage}
    \hspace{1mm}
    \begin{minipage}[b][][b]{0.18\textwidth}
    \centering
        \includegraphics[width=1\linewidth]{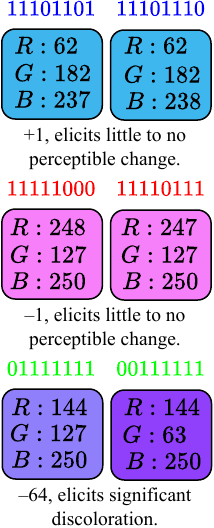}
        \setlength{\belowcaptionskip}{-15pt}
        \caption{Imperceptibility of LSB substitution.}
        \label{fig:fig7}
        \end{minipage}
\end{figure*}
Conventional image steganographic methodologies are constrained by an empirically established payload threshold of approximately $0.4$ bits per pixel \cite{pevny2010using}. Exceeding this critical threshold invariably induces quantifiable distortions, such as deviations in histogram distributions or localized pixel correlation anomalies, which manifest as statistical anomalies detectable by adversarial steganalysis tools and, in severe cases, perceptible to the human eye. However, a new wave of image steganography techniques has emerged with the rapid development of deep learning technologies over the past decade \cite{hayes2017generating,baluja2017hiding,zhu2018hidden}. These modern methods adopt a bifurcated strategy: either by assimilating neural network architectures to refine established algorithmic frameworks---such as leveraging deep learning to discern optimal loci for data concealment---or function as end-to-end image processing pipelines that integrate a carrier image and secret payload into a steganographic construct.
Contemporary deep learning-based methods exhibit certain limitations when juxtaposed with traditional techniques. For instance, they often impose dimension-specific constraints, \textit{e.g.} the requirement for $32 \times 32$ pixel cover images in the method proposed by \citet{hayes2017generating}. Moreover, these methods primarily focus on embedding images within other images, instead of arbitrary messages or binary data streams while not fully probing into security, noise immunity, and finding the upper bounds of quantifiable information that can be effectively concealed. To avoid the temporal overhead of such resource-hungry approaches \cite{zhang2019steganogan}, in this paper, we resort to a rudimentary approach, which is LSB Steganography \cite{neeta2006implementation}. We also introduce the idea of amalgamating cryptography, more specifically Fernet Symmetric Encryption \cite{delfs2007symmetric}, with image steganography to fortify the confidentiality and integrity of the secret message as it traverses the communication channel between the originating and receiving entities. This channel, however, may be susceptible to bit-corruption attacks, which is why we also include the Reed--Solomon error correction codes \cite{reed1960polynomial} of the secret message as a portion of the payload. The harmonious integration of the aforementioned methods culminates in our proposed novel framework \screedsolo, a secure and robust image steganography method where the secret message is Randomly \textbf{\underline{S}}huffled, encrypted using the Fernet Symmetric \textbf{\underline{C}}ipher, and safeguarded using \textbf{\underline{Reed}}--\textbf{\underline{Solo}}mon codes.

%% file: sections/2_Problem_Formulation.tex
\section{Methodology}
The purpose of image steganography is to conceal a secret object that can be either an image $f(x,y)$, or an audio signal $f(t)$, or a piece of text $S$, implicitly within a cover image $g(x,y)$. After the image steganography process is completed, the cover image $g(x,y)$ is called a stego-image $g'(x,y)$, which is then transmitted to the receiver side. If the stego-image reaches the receiver side in its unaltered and uncorrupted form, then the receiver can perform the exact reverse operations to extract the payload secret object from the stego-image $g'(x,y)$.
The undergirding idea beneath steganography is the exploitation of the fact that the eyes of human beings cannot perceive minuscule changes in color luminance (see Figure \ref{fig:fig7}). Figure \ref{fig:fig6} shows an overview of our proposed framework.

%% file: sections/4_Methodology.tex
\subsection{Signal-to-Text Conversion}
For image signals, the first step is to flatten the $c \times M \times N$
image to a 1D array of
$cMN$ pixel values.
If the cover image doesn't have the necessary capacity to harbor the secret payload, then we can opt to perform $k$-bit quantization on the pixel values of the secret image.
The stipulation here is that the $k$ least significant bits of the pixel values may be worth sacrificing if those bits don't contain useful spatial information. For the flattened image $f$, we perform the quantization as follows,
\vspace{-5pt}
\begin{align}
\small
    f^Q = (f \texttt{ >> } (8-k)) + 64
\end{align}
It is to be noted that we add a constant value of $64$ so that the resultant values can be interpreted as ASCII values (\textit{i.e.} for $k$-bit quantization, we end up with values from $64$ to $64 + 2^{k} - 1$). Once this quantization is done, the secret image cannot be extracted in its original form, since quantization is irreversible. After the optional $k$-bit quantization step, we take the ASCII characters and append them to a single text string $S$.

Due to the repetitive nature of audio signals, it is a salient approach to apply signal compression techniques to reduce the number of bits that we have to eventually embed in the cover image. In \screedsolos we use the proprietary \textsc{Deflate} algorithm \cite{oswal2016deflate} of the \texttt{zlib}\footnote{\texttt{\url{https://docs.python.org/3/library/zlib.html}}} package. It is a combination of two lossless compression techniques, namely Huffman coding \cite{huffman1952method} and LZ77 coding \cite{ziv1977universal}.
{\small
\begin{align}
    f^C = \operatorname{DEFLATE}(f(t))
\end{align}
}%
Then we convert the compressed byte stream to its corresponding base-$64$ positional notation, denoted using $(f^C)_{64}$. Since the symbols used in the base-$64$ notation constitute alphanumeric characters, we can simply typecast it as a string.
{\small
\begin{align}
    S = \operatorname{\texttt{str}}((f^C)_{64})
\end{align}
}%
\subsection{\screedsolos Text Encoding}
The sender and receiver sides must share a password $p$ that they use to encode and decode the secret message, respectively.

\subsubsection{Pseudo-random Shuffling}
In the encoding process, at first, we take the message string $S$ and randomly shuffle it. The caveat here is that the shuffling is not absolutely random, but based on a pseudo-random permutation of the string indices. The permutation is uniform, \textit{i.e.}, each of the $|S|!$ possible permutations is equally likely. The seed value $p^s$ that we use for generating the pseudo-random permutation is the hash value of $p$ that is yielded by using the Secure Hash Algorithm-$256$ \cite{penard2008secure}, colloquially known as SHA-$256$.
\begin{align}
    p^s = \operatorname{SHA256}(p)\\
    S' = \operatorname{Shuffle}(S, p^s)
\end{align}
\subsubsection{Fernet Symmetric Encryption}
We generate the cipher key $(p^m)_{64}$ by applying the Message Digest-$5$ hashing algorithm, also known as the MD-$5$ algorithm;
then the numerical hash-value is converted to its corresponding base-$64$ notation.
{\small
\begin{align}
    p^m = \operatorname{MD5}(p)\\
    F = \operatorname{Fernet}(S', (p^m)_{64})
\end{align}
}%
In symmetric encryption algorithms, the same key is used for both encryption and decryption. Fernet
is a cryptographic technique that offers a straightforward method for both authentication and encryption. It utilizes HMAC (Hash-based Message Authentication Code) with SHA-$256$ for authentication and employs symmetric AES-$128$ (Advanced Encryption Standard-$128$) encryption in Cipher Block Chaining (CBC) mode, with PKCS7 (Public Key Cryptography Standards \#7) padding.

\subsubsection{Reed--Solomon Coding}
Reed--Solomon error-correcting codes belong to the family of linear block codes based on the working principle: for an input message
of length $k$, the encoding procedure produces an output codeword of length $n$ where $n \geq k$, guaranteeing the correction of up to {\footnotesize $\displaystyle\floor*{\frac{n-k}{2}}$} errors \cite{reed1960polynomial}. This error-resilience property emerges from the codes' mathematical structure, which leverages the benefit of Theorem \ref{thm:thm1}, by treating message symbols as coefficients in a polynomial equation.
\begin{thm}[Polynomial Uniqueness]
\label{thm:thm1}
    A polynomial $P_{d}(x) = a_0 + a_1x + a_2x^2+ \dots + a_dx^d$ of degree $d\leq n$ that passes through $n+1$ data points $(x_0,y_0),\dots,(x_n,y_n)$ is unique.  
\end{thm}
By sampling this polynomial at multiple points to generate redundant symbols, Reed–Solomon encoding creates a system where the original polynomial---and thus the original message---can be reconstructed even when some sample points become corrupted. Therefore, given a steganography algorithm which, on average, returns an incorrect bit with probability $p$, it is desirable that,
{\small
\begin{align}
\label{eq:reedsolo}
    \mathds{E}[X= \text{\# of corrupted bits}] = p \times n \leq \frac{n-k}{2}
\end{align}
}%
The effective information throughput, signified by the ratio $\frac{k}{n}$, quantifies the average number of \textit{`real'} data bits conveyed per \textit{`message'} data bit.
The main pitfall of resorting to Reed--Solomon coding in our framework is the aforementioned additional data overhead of the error-correction codes, \textit{i.e.}, the size of the encoded message $R$ will be higher than the size of the Fernet encrypted message $F$. The advantage, however, is of course, the ability of the framework to withstand the bit-corruption of at most {\footnotesize $\floor*{\frac{n-k}{2}}$} bits.
{\small
\begin{align}
    R = \operatorname{ReedSolo}(F)
\end{align}
}%
\subsubsection{Forming the Binary Message}
$R$ consists of characters that have ASCII values $\in [0, 127]$, and each of these characters take at most $7$ bits to be expressed in the binary notation. We replace the characters of $R$ with their corresponding base-$2$ values, thus forming $(R)_2$. For the decoding process at the receiver side, it is necessary to know the length of this entire binary message. So we prepend this length $L$ using a bitset of $32$ bits (\textit{i.e.} $(L)_2$) at the anterior part of the binary message.
\vspace{-2mm}
{\small
\begin{align}
    (R)_2 = \operatorname{\texttt{String2Binary}}(R)\\
    L = \operatorname{Length}((R)_2)\\
    (L)_2 = \operatorname{\texttt{Decimal2Binary}}(L)\\
    M = (L)_2 \concat (R)_2
\end{align}
}%
We posit that the final binary message $M$ is the concatenated binary string $(L)_2 \concat (R)_2$.

\subsubsection{LSB Steganography}
LSB (Least Significant Bit) Steganography is a technique in which secret information is embedded into the least significant bits of pixels in a digital image. By modifying only the last few bits of each pixel value, LSB steganography exploits the fact that such small changes are imperceptible to the human eye (as evident in Figure \ref{fig:fig7}), allowing the secret message to remain hidden while preserving the appearance of the image.
In a typical $n$-bit image, each pixel's color value can be represented as an $n$-bit binary number,
and the least significant bit of each pixel can be used to encode the secret data. 
{\small
\begin{align}
\label{eq:lsb}
    g'(x,y) = \begin{cases}
        g(x,y) \oplus 1;\quad \text{if pixel's LSB $\neq$ message's bit}\\
        g(x,y);\quad\quad \hspace{2.75mm} \text{otherwise}
    \end{cases}
\end{align}
}%
The capacity of LSB steganography depends on the number of bits available in the cover image. In an $8$-bit grayscale image, each pixel can store $1$ bit of the message. Thus, for an image with $M \times N$ pixels, the total message capacity is $M \times N$ bits. For a $24$-bit color image, each pixel can store $3$ bits (one in each of the 3 channels), allowing a higher embedding capacity $3 \times M \times N$ bits. As per Equation \ref{eq:lsb}, we simply alter the LSB of the cover image $g(x,y)$ if the pixel in question has an LSB that is different from the bit that we are trying to embed. Otherwise, we keep the pixel unaltered. The resultant image $g'(x,y)$ is referred to as the stego-image.

%% file: sections/5_Experiment.tex
\section{Experiment}
\label{sec:experiment}
We experiment with multiple modalities of secret objects (\textit{e.g.} text, audio, and image) and embed them in cover images of different resolutions. The metrics that we use to evaluate the performance of \screedsolos are Cover Image Loss, Cosine Similarity (CSim), Mean Square Error (MSE), Peak Signal-to-Noise Ratio (PSNR), Structural Similarity Index Measure (SSIM), Variation of Information (VI) Hausdorff Distance (HDist), and Normalized Root Mean Square Error (NRMSE).
\begin{figure}[t]
    \centering
    \subfloat[\centering Pillars of Creation (Cover Image).]{\includegraphics[width=0.25\linewidth]{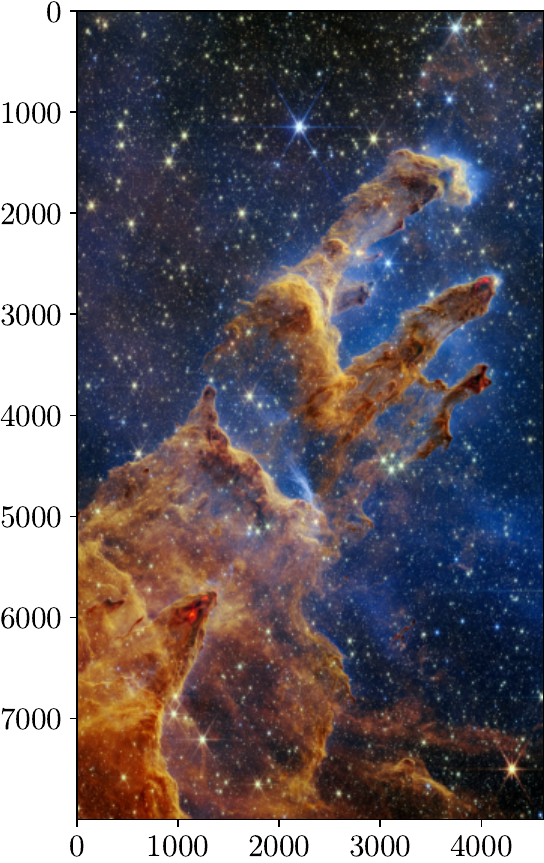}\label{fig:fig9a}}
    \subfloat[\centering Pillars of Creation (Stego-image).]{\includegraphics[width=0.25\linewidth]{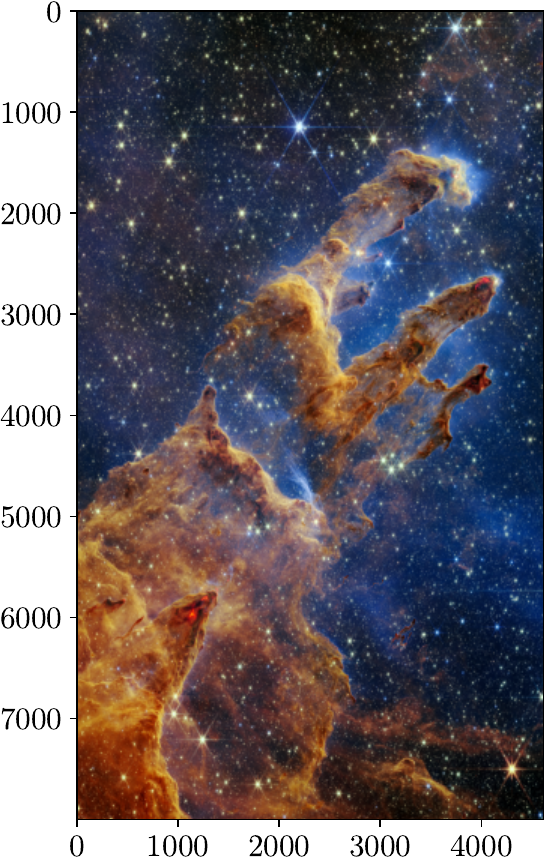}\label{fig:fig9b}}
    \subfloat[\centering Difference Image.]{\includegraphics[width=0.25\linewidth]{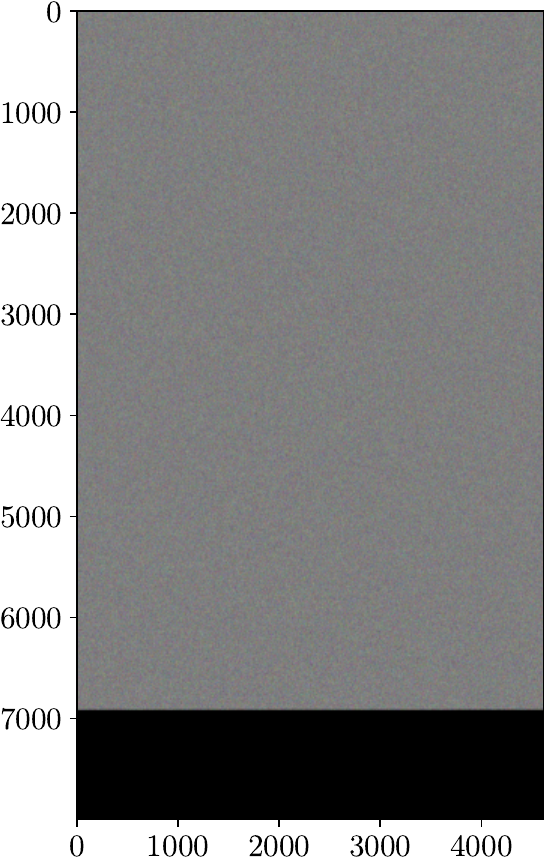}\label{fig:fig9c}}
    \setlength{\belowcaptionskip}{-18pt}
    \caption{Hiding \textit{The Complete Works of Shakespeare} inside the \textit{Pillars of Creation}.}
    \label{fig:fig9}
\end{figure}
\vspace{-6.5mm}
\subsection{Text Hiding}
\subsubsection{Text Corpus}
In order to test out the secret text embedding process, we consider the classical literature corpus called \textit{The Complete Works of Shakespeare}
obtained from Project Gutenberg.\footnote{\texttt{\url{https://www.gutenberg.org/ebooks/100}}} \textit{The Complete Works of William Shakespeare} is the standard name given to any volume containing all the plays and poems of William Shakespeare. The entire text body has a length of 5,458,195 characters. After compression, the size shrinks down to 2,680,939 characters. After the encryption process, the length becomes 11,972,988. After conversion to binary notation, the total bits to embed becomes 95,783,904.

\subsubsection{Cover Image}
The cover image that we consider for hiding this payload is a Near-Infrared Camera (NIRCam) image taken by the James Webb Space Telescope in 2022 (see Figure \ref{fig:fig9a}). The resolution of the image is $14589 \times 8423 \times 3$.
\begin{table}[H]
\centering
\small
\setlength\tabcolsep{1.5pt}
\begin{tabular}{|c|c|c|c|c|c|c|c|}
\hline
\textbf{Loss} & \textbf{CSim} & \textbf{MSE} & \textbf{PSNR} & \textbf{SSIM} & \textbf{VI} & \textbf{HDist} & \textbf{NRMSE} \\ \hline
$86.422$ & $0.999$ & $1.296$ & $51.774$ & $0.996$ & $\langle 0.982, 0.982 \rangle$ & $3.464$ & $0.007$ \\ \hline
\end{tabular}
\setlength{\belowcaptionskip}{-15pt}
\caption{Quantitative evaluation results for Figure \ref{fig:fig9}.}
\label{tab:tab1}
\end{table}

\subsection{Audio Hiding}
\subsubsection{Audio Signal}
In order to test out the secret audio embedding process, we consider the classical music score called \textit{Moonlight Sonata}. The Piano Sonata No. 14 in C-sharp minor, marked \textit{Quasi una fantasia}, Op. 27, No. 2, is a piano sonata by Ludwig van Beethoven.
The duration of the score is 15 minutes, and due to the repetitive nature of the score, we compress it using \textsc{Deflate}. In its original state, the audio file has a length of 61,777,387 bytes. After compression, it becomes 35,999,084 bytes in length. We convert the audio to text by maintaining the pipeline outlined in Figure \ref{fig:fig6}. The encrypted text has a length of 2,23,727,460 and the binary message is of length 1,789,819,680 bits.
\begin{figure}[t]
    \centering
     \subfloat[\centering Moonlight Sonata (Secret).]{\includegraphics[width=0.25\linewidth]{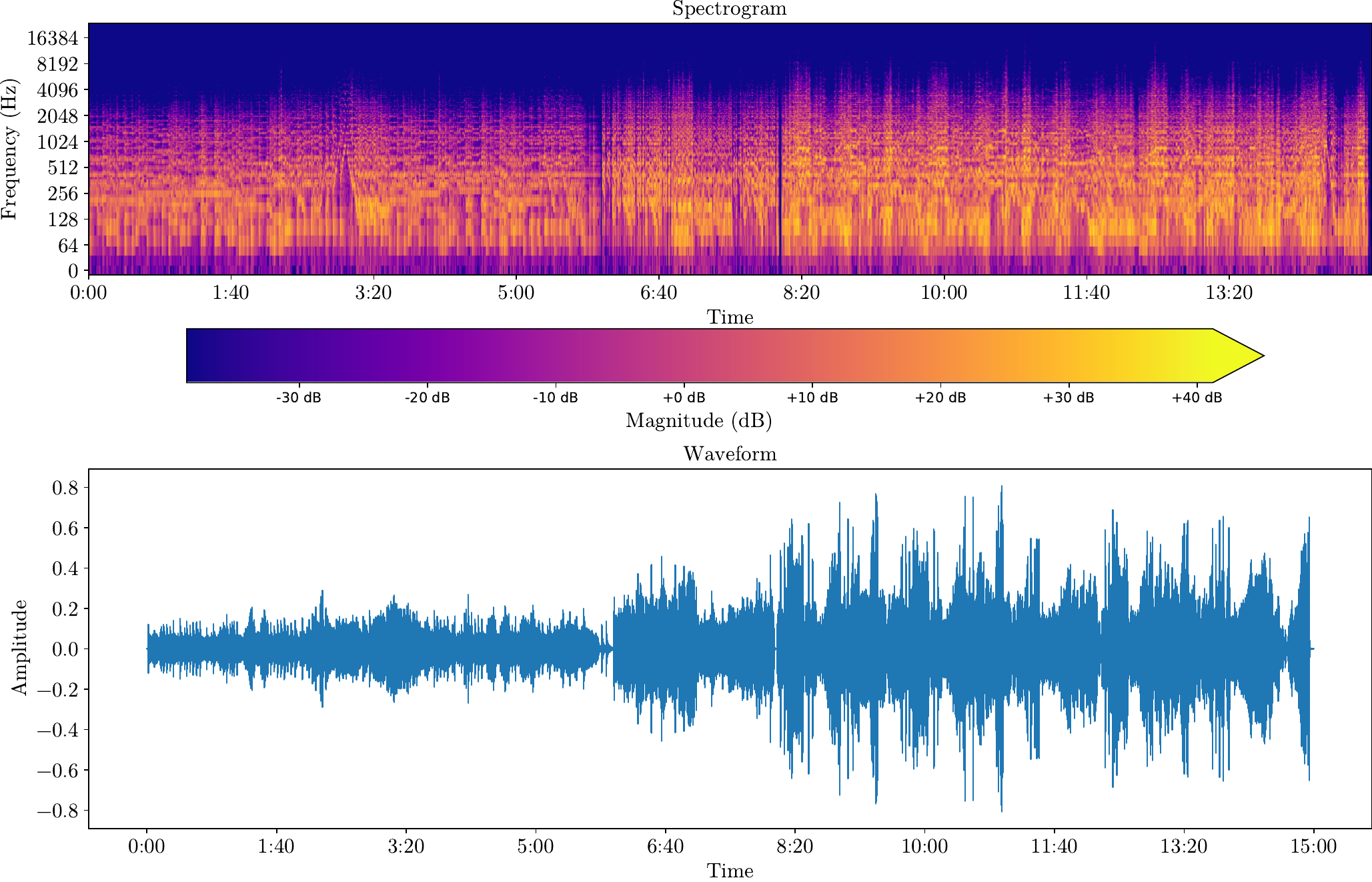}\label{fig:fig10a}}
    \subfloat[\centering The Moon (Cover Image).]{\includegraphics[width=0.25\linewidth]{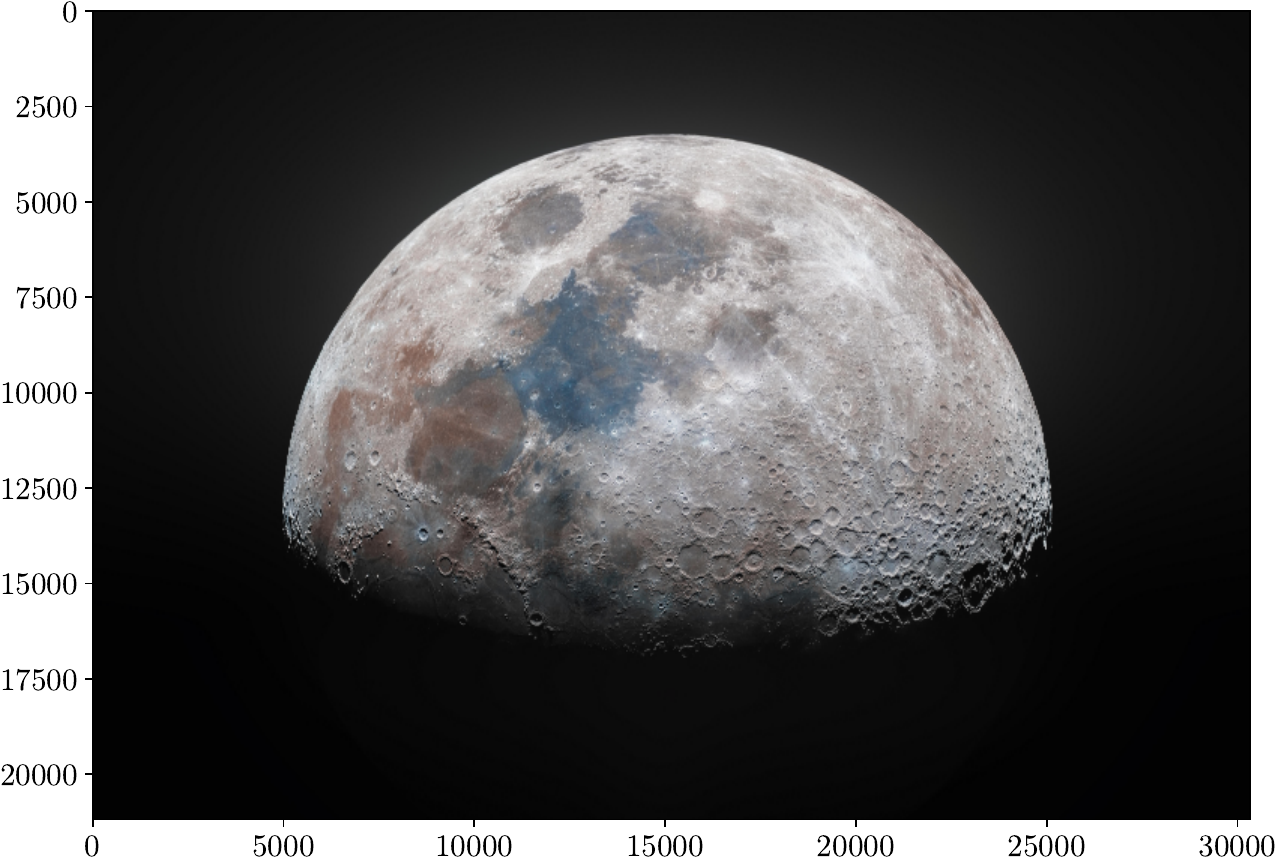}\label{fig:fig10b}}
    \subfloat[\centering The Moon (Stego-image).]{\includegraphics[width=0.25\linewidth]{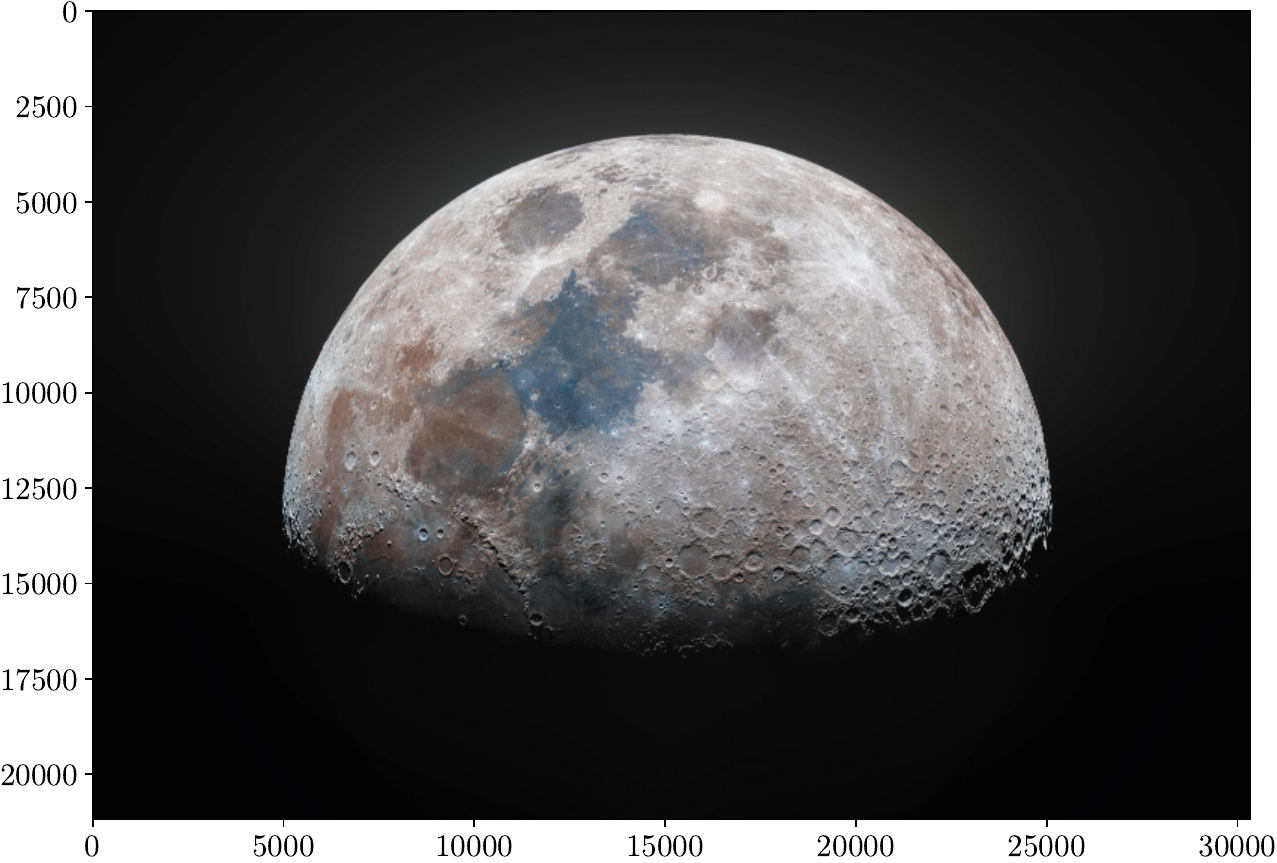}\label{fig:fig10c}}
    \subfloat[\centering Difference Image.]{\includegraphics[width=0.25\linewidth]{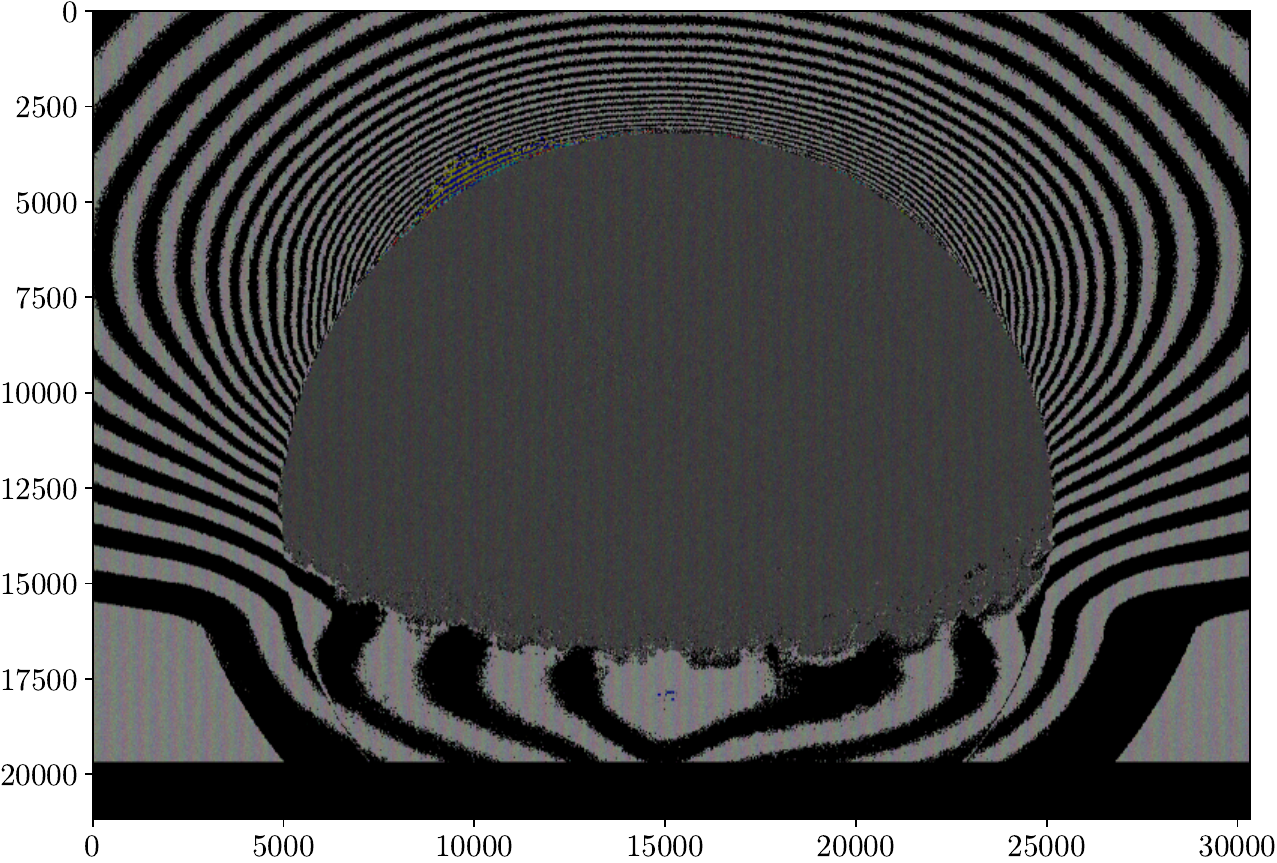}\label{fig:fig10d}}
    \setlength{\belowcaptionskip}{-15pt}
    \caption{Hiding the \textit{Moonlight Sonata} inside the \textit{Moon}.}
    \label{fig:fig10}
\end{figure}
\subsubsection{Cover Image}
The cover image that we consider for hiding this payload is a detailed image of the Moon captured by astrophotographer Andrew McCarthy (see Figure \ref{fig:fig10b}). 
The resolution of the image is $21188 \times 30328 \times 3$.
\vspace{-2mm}
\begin{table}[H]
\centering
\small
\setlength\tabcolsep{1.5pt}
\begin{tabular}{|c|c|c|c|c|c|c|c|}
\hline
\textbf{Loss} & \textbf{CSim} & \textbf{MSE} & \textbf{PSNR} & \textbf{SSIM} & \textbf{VI} & \textbf{HDist} & \textbf{NRMSE} \\ \hline
$92.844$ & $0.999$ & $1.412$ & $57.653$ & $0.936$ & $\langle 0.902, 0.828 \rangle$ & $2.0$ & $0.007$ \\ \hline
\end{tabular}
\setlength{\belowcaptionskip}{-15pt}
\caption{Quantitative evaluation results for Figure \ref{fig:fig10}.}
\label{tab:tab2}
\end{table}
\subsection{Image Hiding}
\vspace{-1.5mm}
\subsubsection{Image Payload}
We use the Chalk image shown in Figure \ref{fig:fig11b} as our secret image. We use $k = 5$ while performing the optional $k$-bit quantization. As a pragmatic pursuit, we downsample the image to the resolution $100 \times 100 \times 3$, so the total number of pixel values that we encode is 30,000.

\subsubsection{Cover Image}
We use the Lenna image shown in Figure \ref{fig:fig11a} as our cover image. The image's resolution is $512 \times 512$.
\vspace{-2mm}
\begin{table}[H]
\centering
\small
\setlength\tabcolsep{1.5pt}
\begin{tabular}{|c|c|c|c|c|c|c|c|}
\hline
\textbf{Loss} & \textbf{CSim} & \textbf{MSE} & \textbf{PSNR} & \textbf{SSIM} & \textbf{VI} & \textbf{HDist} & \textbf{NRMSE} \\ \hline
$67.142$ & $0.999$ & $1.003$ & $52.888$ & $0.996$ & $\langle 0.914, 0.878 \rangle$ & $2.0$ & $0.006$ \\ \hline
\end{tabular}
\setlength{\belowcaptionskip}{-13pt}
\caption{Quantitative evaluation results for Figure \ref{fig:fig11}.}
\label{tab:tab3}
\end{table}
\begin{figure}[t]
    \centering
    \subfloat[\centering Lenna (Cover).]{\includegraphics[width=0.2\linewidth]{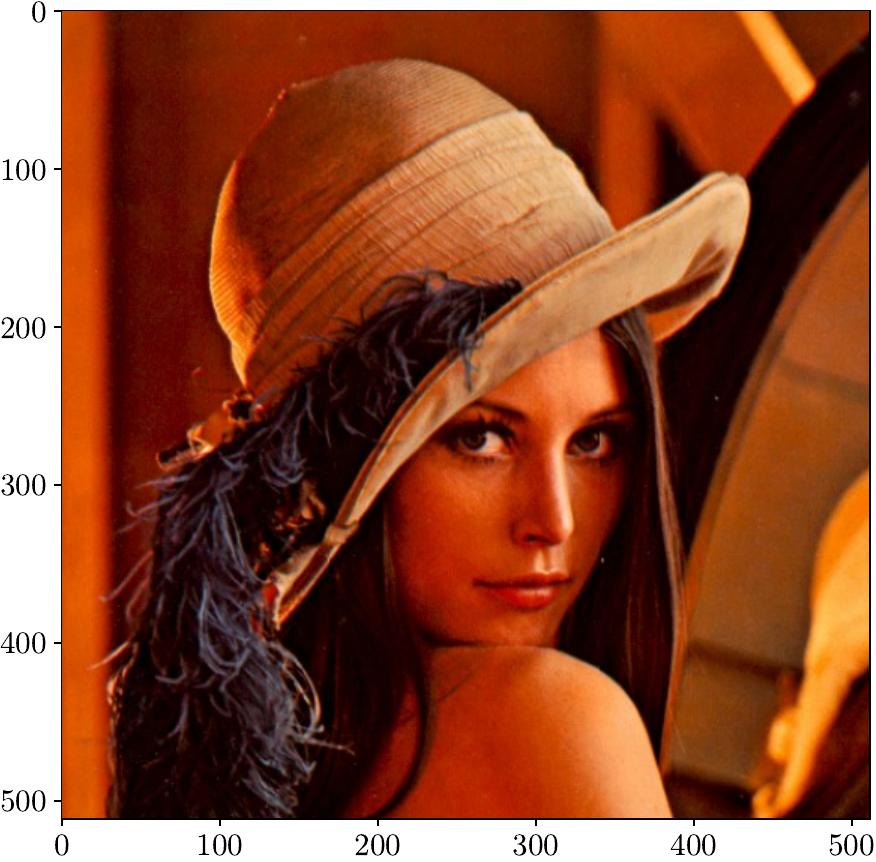}\label{fig:fig11a}}
    \subfloat[\centering Chalk (Secret).]{\includegraphics[width=0.2\linewidth]{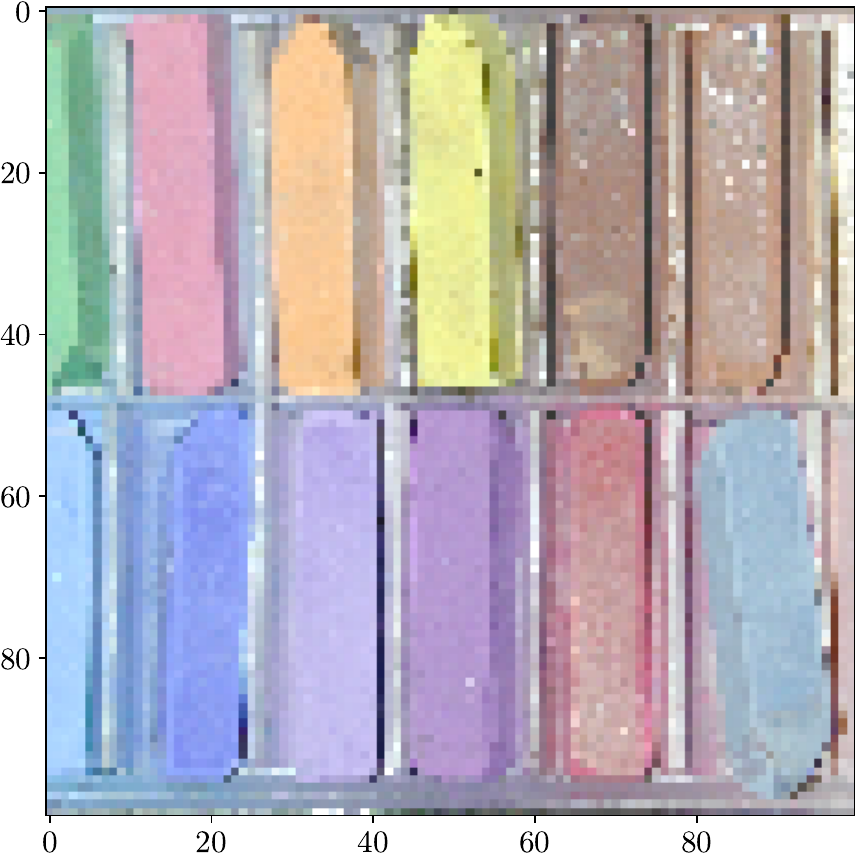}\label{fig:fig11b}}
    \subfloat[\centering Lenna (Stego-image).]{\includegraphics[width=0.2\linewidth]{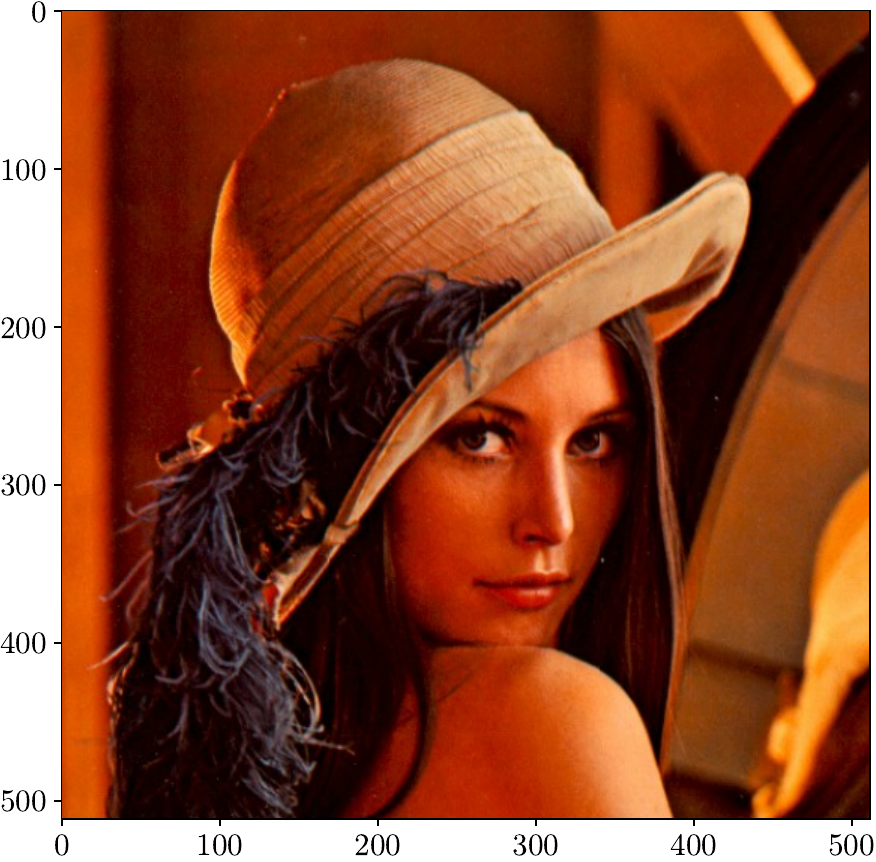}\label{fig:fig11c}}
    \subfloat[\centering Difference Image.]{\includegraphics[width=0.2\linewidth]{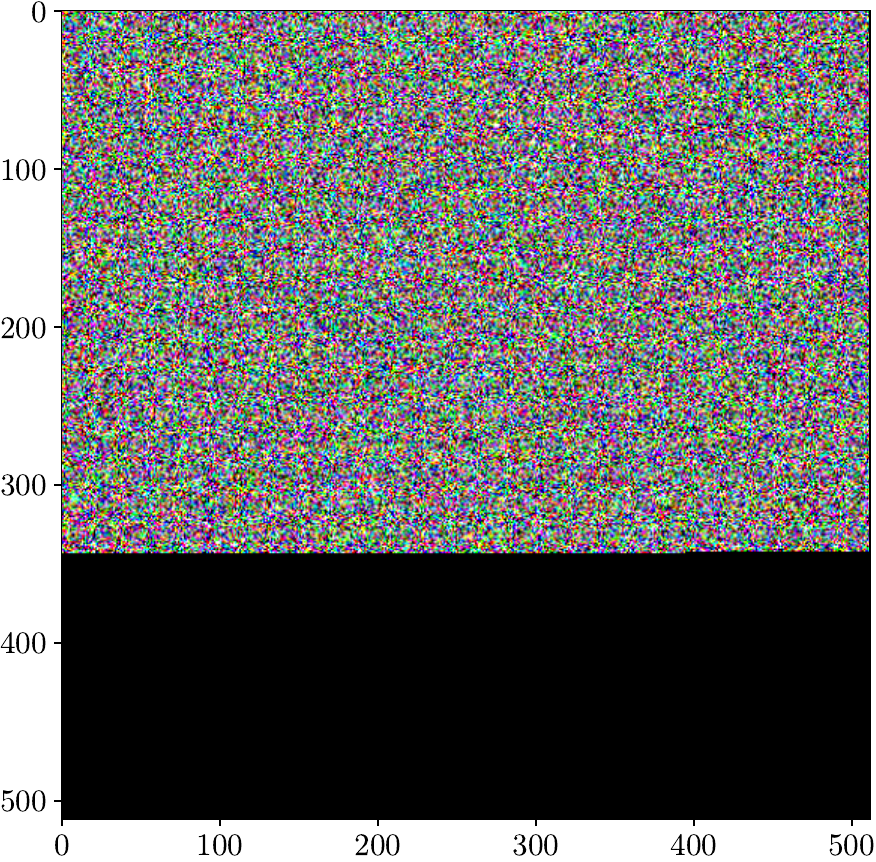}\label{fig:fig11d}}
    \subfloat[\centering Recovered Image {\scriptsize ($k=5$)}.]{\includegraphics[width=0.2\linewidth]{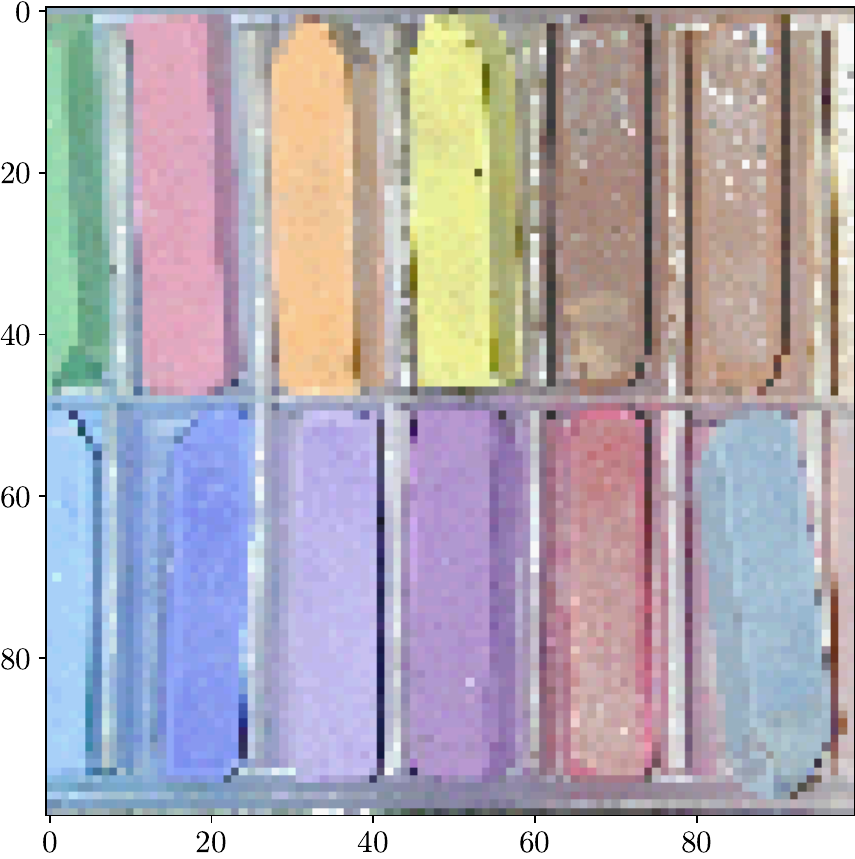}\label{fig:fig11e}}
    \\
    \vspace{-2mm}
    \subfloat[\centering Histogram of Cover Image\\(Figure \ref{fig:fig11a}).]{\includegraphics[width=0.5\linewidth]{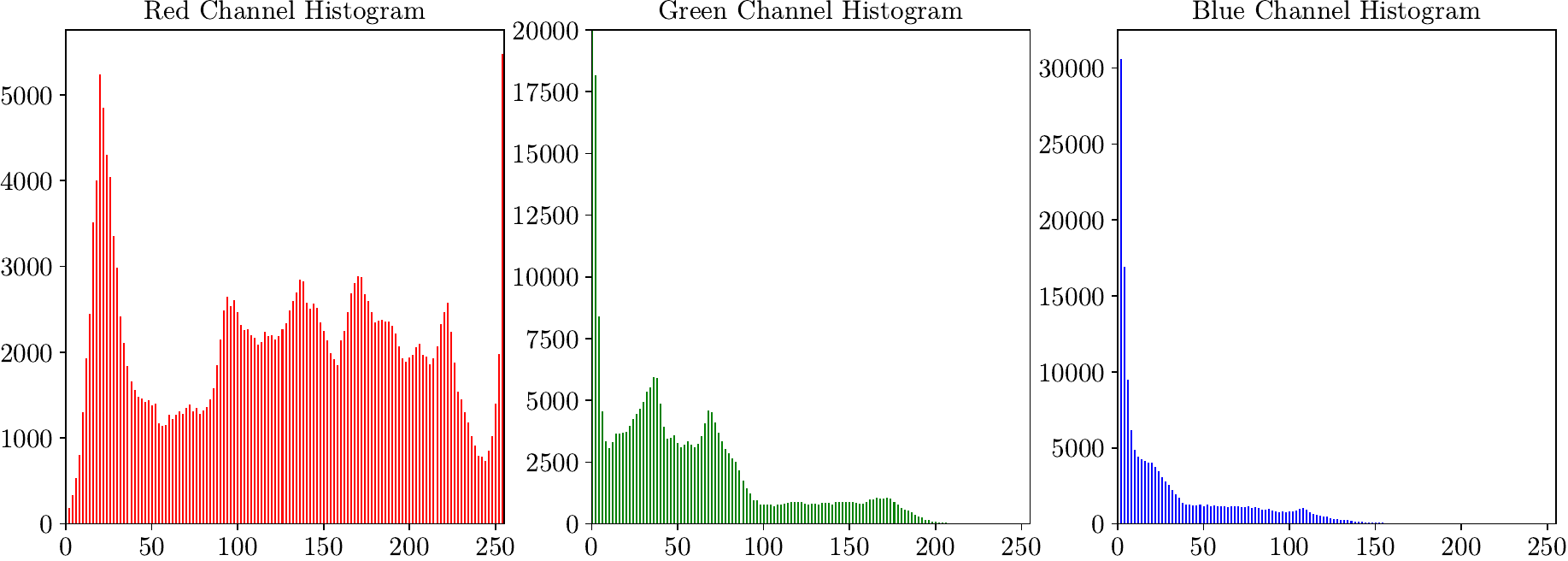}\label{fig:fig11f}}
    \subfloat[\centering Histogram of Stego-image\\(Figure \ref{fig:fig11c}).]{\includegraphics[width=0.5\linewidth]{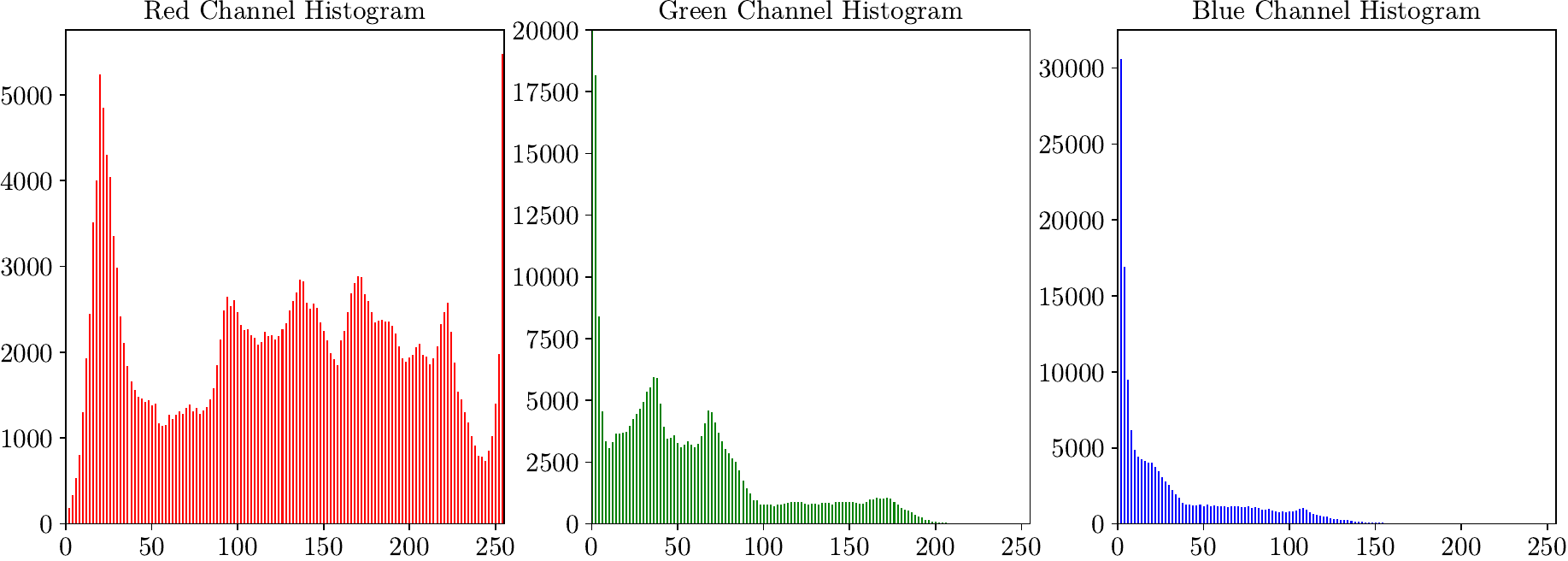}\label{fig:fig11g}}
    \setlength{\belowcaptionskip}{-15pt}
    \caption{Hiding the \textit{Chalk} image inside the \textit{Lenna} image.}
    \label{fig:fig11}
\end{figure}
\subsubsection{Steganalysis Attacks --- Noise and Bit-corruption}
\begin{figure}[t]
    \centering
    \subfloat[\centering Salt \& Pepper ({\footnotesize$p_{s},p_{p}\hspace{-1pt}=\hspace{-1pt}3\operatorname{e}^{-2}$})]{\includegraphics[width=0.25\linewidth]{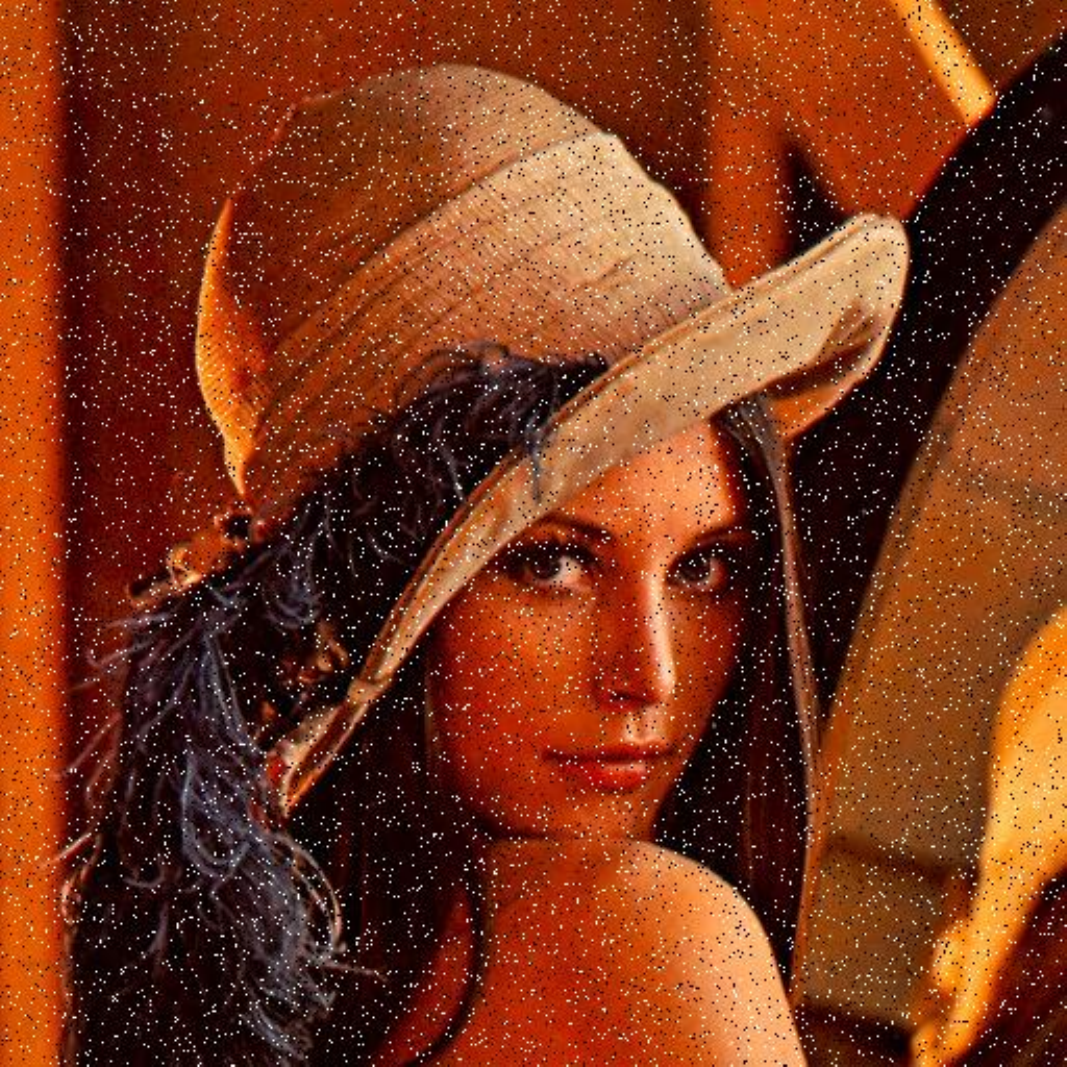}\label{fig:fig12a}}
    \subfloat[\centering Gaussian Noise ({\footnotesize$\mu\hspace{-1pt}=\hspace{-1pt}0, \sigma\hspace{-1pt}=\hspace{-1pt}0.63$})]{\includegraphics[width=0.25\linewidth]{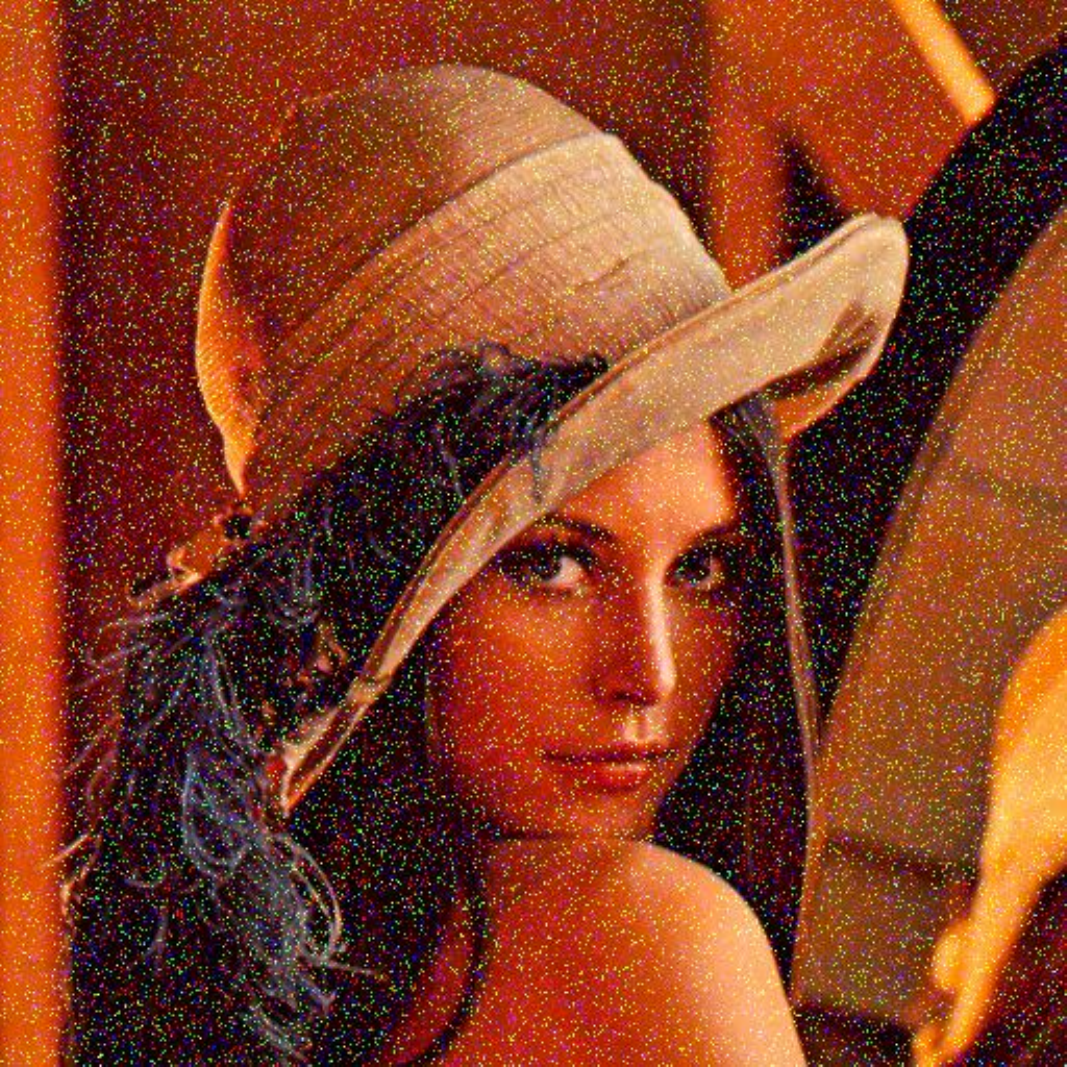}\label{fig:fig12b}}
    \subfloat[\centering Speckle Noise ({\footnotesize$\mu\hspace{-1pt}=\hspace{-1pt}0, \sigma\hspace{-1pt}=\hspace{-1pt}1\operatorname{e}^{-1}$})]{\includegraphics[width=0.25\linewidth]{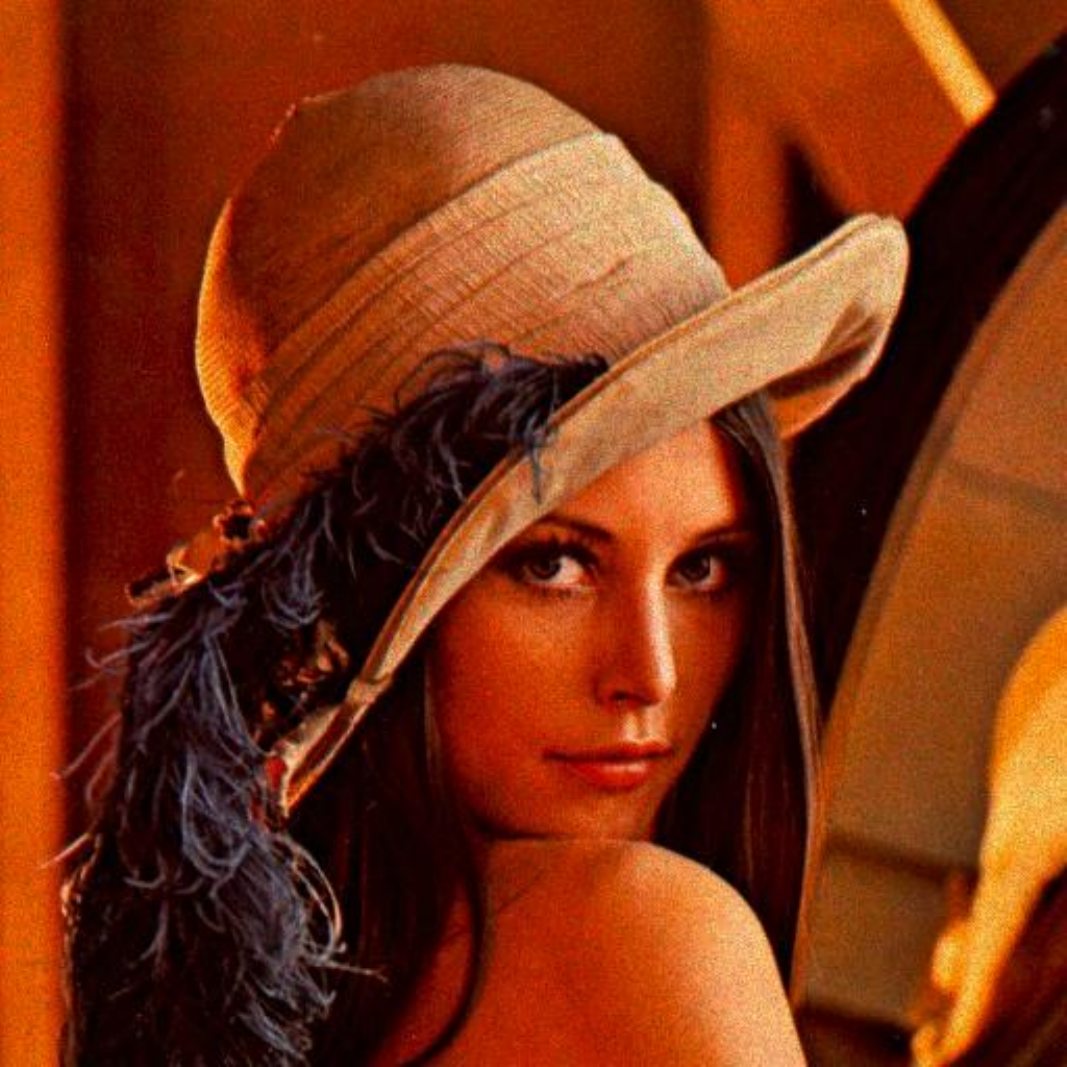}\label{fig:fig12c}}
    \subfloat[\centering Poisson Noise ({\footnotesize$\lambda\hspace{-1pt}=\hspace{-1pt}9\operatorname{e}^{-1}$})]{\includegraphics[width=0.25\linewidth]{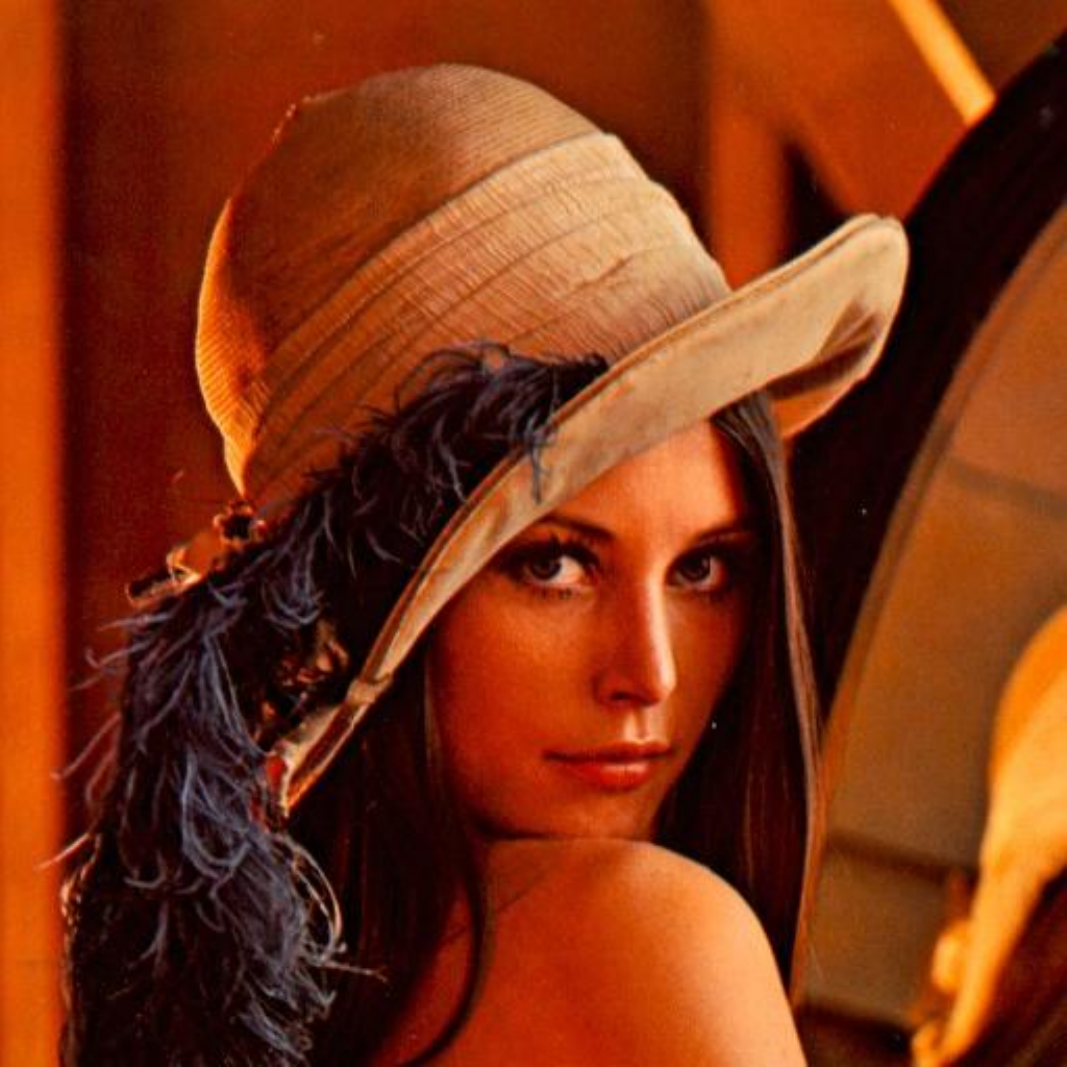}\label{fig:fig12d}}
    \setlength{\belowcaptionskip}{-15pt}
    \caption{Applying different types of noise on the Stego-image.}
    \label{fig:fig12}
\end{figure}

The incorporation of Reed--Solomon error-correction codes enables the stego-image to withstand corruption and noise contingent on the fact that the noise must not alter more than half of the number of error-correction bits. We test this proposition with 4 types of noise (see Figure \ref{fig:fig12}). In tandem, Table \ref{tab:tab4} shows \screedsolo's resilience against visual/spatial perturbations.
\vspace{-2mm}
\begin{table}[H]
\centering
\small
\setlength\tabcolsep{1.5pt}
\begin{tabular}{|c|c|c|c|c|c|c|c|}
\hline
\textbf{Noise} & \textbf{CSim} & \textbf{MSE} & \textbf{PSNR} & \textbf{SSIM} & \textbf{VI} & \textbf{HDist} & \textbf{NRMSE} \\ \hline
Figure \ref{fig:fig12a} & $0.976$ & $1349.961$ & $21.598$ & $0.665$ & $0.315$ & $2.0$ & $0.221$ \\ \hline
Figure \ref{fig:fig12b} & $0.979$ & $1188.227$ & $22.153$ & $0.725$ & $0.317$ & $2.0$ & $0.207$ \\ \hline
Figure \ref{fig:fig12c} & $0.999$ & $1.340$ & $51.628$ & $0.995$ & $1.855$ & $2.0$ & $0.006$ \\ \hline
Figure \ref{fig:fig12d} & $0.999$ & $0.026$ & $68.604$ & $0.999$ & $0.143$ & $2.0$ & $0.0009$ \\ \hline
\end{tabular}
\setlength{\belowcaptionskip}{-13pt}
\caption{Effect of noise on the stego-image (Figure \ref{fig:fig11c}).}
\label{tab:tab4}
\end{table}
\begin{prop}[Parity Consistency]
    For a given $3\times M \times N$ LSB stego-image $f(x,y)$ harboring $k$ message bits and $(n-k)$ Reed–Solomon error correction bits, and steganalysis transformation $T$, if $S$ is the set of $\forall x, \forall y, \langle x, y\rangle \in \{ \langle x, y\rangle \mid T(f(x, y)) \land 1 = f(x, y) \land 1 \}$, then a necessary but not sufficient condition for a successful transmission is $|S|\geq \ceil*{\frac{n+k}{2}}$.
\end{prop}
Owing to this insufficiency, we mathematically analyze the noise immunity likelihood for \screedsolo's stego-images.
\begin{thm}[Survival Probability]
    If the random variable $X$ denotes the number of uncorrupted least significant bits for a given steganalysis transformation $T$ on a  $3\times M \times N$ stego-image $f(x,y)$ harboring $k$ message bits and $(n-k)$ Reed–Solomon error correction bits such that $n \leq 3MN$, then 
    the probability of successful payload transmission is
    \vspace{-2mm}
    {\small
    \begin{align*}
    \mathds{P}\left(X\geq \ceil*{\frac{n+k}{2}}\middle|f,T\right)=\displaystyle\sum_{i=\ceil*{\frac{n+k}{2}}}^{n}\binom{n}{i} \times \frac{\binom{3MN}{n}}{2^{n}}
    \end{align*}
    }%
    \vspace{-15pt}
    \proof Each channel of each pixel constitutes only $1$ bit (LSB) of information related to the secret message. So, for $m$-bit pixel channels, $\mathds{P}(\text{LSB remains unchanged}) = \frac{2^{m-1}}{2^m} = \frac{1}{2}$. This implies that $X$ follows the binomial distribution, \textit{i.e.}, $X \sim \operatorname{Bin}(n,p=0.5)$. So, following Equation \ref{eq:reedsolo}, the CDF $F(\ceil*{\frac{n+k}{2}};n,p)$ can be obtained as follows
    {\footnotesize
    \begin{align*}
    \mathds{P}\left(X\geq \ceil*{\frac{n+k}{2}}\right)&=\left[\frac{\displaystyle\binom{n}{\ceil*{\frac{n+k}{2}}}}{2^n}+\frac{\displaystyle\binom{n}{\ceil*{\frac{n+k}{2}}+1}}{2^n}+\dots+\frac{\displaystyle\binom{n}{n}}{2^n}\right]\\
    \times \binom{3MN}{n}&=\displaystyle\sum_{i=\ceil*{\frac{n+k}{2}}}^{n}\binom{n}{i} \times \frac{\binom{3MN}{n}}{2^{n}}\quad\square\text{ Q.E.D.}
    \end{align*}
    }%
\end{thm}
\subsubsection{Steganalysis Tools}
We use the \texttt{aletheia} toolbox\footnote{\texttt{\url{https://github.com/daniellerch/aletheia}}}
developed by \citet{Aletheia}. The framework performs well across multiple passive steganalysis methods.
\subsection{Result Discussion}
The quantitative evaluations of the steganographic outputs, as presented in Tables \ref{tab:tab1}, \ref{tab:tab2}, and \ref{tab:tab3} demonstrate the efficacy of the proposed methodology in achieving nigh-imperceptible data embedding across different cover images. The cover image loss, of course, is proportional to the size of the secret message payload and can be pictorially visualized from the difference images in Figures \ref{fig:fig9c}, \ref{fig:fig10d}, and \ref{fig:fig11d}. We also observe high visual fidelity between the original cover images and their corresponding stego-images, as is evidenced by the histogram comparison (Figures \ref{fig:fig11f}--\ref{fig:fig11g}) and values for the other image similarity and quality metrics in the aforementioned tables.

%% file: sections/6_Conclusion_and_Future_Work.tex
\section{Conclusion and Future Work}
In this paper, we present \screedsolo, a novel framework for image steganography
that offers a secure and corruption-resilient method for embedding arbitrary binary data into images, achieving a high payload capacity of $3$ bits per pixel with minimal spatial perturbations and stochastically effective obfuscated transmission.
There are several avenues for future work to potentiate this framework. First, further optimizations could be applied to improve the embedding capacity, particularly for applications requiring higher payloads without sacrificing security. Additionally, although the framework proves resistant to simple active steganalysis attacks, more advanced and adversarial steganalysis methods could be explored to ensure its defense against increasingly sophisticated attacks.